\pdfoutput=1
\documentclass[12pt,a4paper]{article}

\usepackage{ifthen} 
\newboolean{pdflatex}
\setboolean{pdflatex}{true} 

\newboolean{articletitles}
\setboolean{articletitles}{true} 

\newboolean{uprightparticles}
\setboolean{uprightparticles}{false}

\usepackage{microtype}
\usepackage{lineno}  
\usepackage{xspace}

\usepackage{graphicx}  
\usepackage{color}
\usepackage{colortbl}
\graphicspath{{./figs/}} 
\usepackage{makecell}
\usepackage{array}
\usepackage{multirow}

\usepackage{chngpage}
\usepackage{amsmath} 
\usepackage{amssymb}
\usepackage{amsfonts}
\usepackage{upgreek}

\newcommand*\patchAmsMathEnvironmentForLineno[1]{%
\expandafter\let\csname old#1\expandafter\endcsname\csname #1\endcsname
\expandafter\let\csname oldend#1\expandafter\endcsname\csname
end#1\endcsname
 \renewenvironment{#1}%
   {\linenomath\csname old#1\endcsname}%
   {\csname oldend#1\endcsname\endlinenomath}%
}
\newcommand*\patchBothAmsMathEnvironmentsForLineno[1]{%
  \patchAmsMathEnvironmentForLineno{#1}%
  \patchAmsMathEnvironmentForLineno{#1*}%
}
\AtBeginDocument{%
\patchBothAmsMathEnvironmentsForLineno{equation}%
\patchBothAmsMathEnvironmentsForLineno{align}%
\patchBothAmsMathEnvironmentsForLineno{flalign}%
\patchBothAmsMathEnvironmentsForLineno{alignat}%
\patchBothAmsMathEnvironmentsForLineno{gather}%
\patchBothAmsMathEnvironmentsForLineno{multline}%
}

\usepackage{hyperref}   
\usepackage[all]{hypcap}

\def\lhcb {\mbox{LHCb}\xspace}
\def\ux85 {\mbox{UX85}\xspace}

\ifthenelse{\boolean{uprightparticles}}%
{

 \def\Ppi         {\ensuremath{\uppi}\xspace}

 \def\PDelta      {\ensuremath{\Delta}\xspace}                 
 \def\PXi      {\ensuremath{\Xi}\xspace}                 
 \def\PLambda      {\ensuremath{\Lambda}\xspace}                 
 \def\PSigma      {\ensuremath{\Sigma}\xspace}                 
 \def\POmega      {\ensuremath{\Omega}\xspace}                 
 \def\PUpsilon      {\ensuremath{\Upsilon}\xspace}                 
 
 \def\PB      {\ensuremath{\mathrm{B}}\xspace}                 
                  
 \def\PD      {\ensuremath{\mathrm{D}}\xspace}

 \def\PK      {\ensuremath{\mathrm{K}}\xspace}

 \def\Pb      {\ensuremath{\mathrm{b}}\xspace}                 
 \def\Pc      {\ensuremath{\mathrm{c}}\xspace}

 \def\Pi      {\ensuremath{\mathrm{i}}\xspace}

 \def\Ps      {\ensuremath{\mathrm{s}}\xspace}

}
{

 \def\Ppi         {\ensuremath{\pi}\xspace}

 \mathchardef\PDelta="7101
 \mathchardef\PXi="7104
 \mathchardef\PLambda="7103
 \mathchardef\PSigma="7106
 \mathchardef\POmega="710A
 \mathchardef\PUpsilon="7107
                  
 \def\PB      {\ensuremath{B}\xspace}                 
                  
 \def\PD      {\ensuremath{D}\xspace}

 \def\PK      {\ensuremath{K}\xspace}

 \def\Pb      {\ensuremath{b}\xspace}                 
 \def\Pc      {\ensuremath{c}\xspace}

 \def\Pi      {\ensuremath{i}\xspace}

 \def\Ps      {\ensuremath{s}\xspace}

}

\def\squark    {\ensuremath{\Ps}\xspace}

\def\cquark    {\ensuremath{\Pc}\xspace}
\def\cquarkbar {\ensuremath{\overline \cquark}\xspace}
\def\ccbar     {\ensuremath{\cquark\cquarkbar}\xspace}
\def\bquark    {\ensuremath{\Pb}\xspace}
\def\bquarkbar {\ensuremath{\overline \bquark}\xspace}
\def\bbbar     {\ensuremath{\bquark\bquarkbar}\xspace}

\def\pion  {\ensuremath{\Ppi}\xspace}
\def\piz   {\ensuremath{\pion^0}\xspace}

\def\pip   {\ensuremath{\pion^+}\xspace}
\def\pim   {\ensuremath{\pion^-}\xspace}

\def\kaon  {\ensuremath{\PK}\xspace}
  \def\Kbar  {\kern 0.2em\overline{\kern -0.2em \PK}{}\xspace}

\def\Kz    {\ensuremath{\kaon^0}\xspace}
\def\Kzb   {\ensuremath{\Kbar^0}\xspace}
\def\KzKzb {\ensuremath{\Kz \kern -0.16em \Kzb}\xspace}
\def\Kp    {\ensuremath{\kaon^+}\xspace}
\def\Km    {\ensuremath{\kaon^-}\xspace}

\def\KpKm  {\ensuremath{\Kp \kern -0.16em \Km}\xspace}
\def\KS    {\ensuremath{\kaon^0_{\rm\scriptscriptstyle S}}\xspace} 
\def\KL    {\ensuremath{\kaon^0_{\rm\scriptscriptstyle L}}\xspace}

  \def\Dbar    {\kern 0.2em\overline{\kern -0.2em \PD}{}\xspace}
\def\D       {\ensuremath{\PD}\xspace}

\def\Dz      {\ensuremath{\D^0}\xspace}
\def\Dzb     {\ensuremath{\Dbar^0}\xspace}
\def\DzDzb   {\ensuremath{\Dz {\kern -0.16em \Dzb}}\xspace}
\def\Dp      {\ensuremath{\D^+}\xspace}
\def\Dm      {\ensuremath{\D^-}\xspace}
\def\Dpm     {\ensuremath{\D^\pm}\xspace}

\def\DpDm    {\ensuremath{\Dp {\kern -0.16em \Dm}}\xspace}

\def\Dstarp  {\ensuremath{\D^{*+}}\xspace}

\def\Ds      {\ensuremath{\D^+_\squark}\xspace}
\def\Dsp     {\ensuremath{\D^+_\squark}\xspace}

\def\B       {\ensuremath{\PB}\xspace}
  \def\Bbar    {\kern 0.18em\overline{\kern -0.18em \PB}{}\xspace}

\def\Bz      {\ensuremath{\B^0}\xspace}

  \def\Y#1S{\ensuremath{\PUpsilon{(#1S)}}\xspace}

\def\Lbar {\ensuremath{\kern 0.1em\overline{\kern -0.1em\PLambda}}\xspace}

\def\to                 {\ensuremath{\rightarrow}\xspace}

\def\CP                {\ensuremath{C\!P}\xspace}

\def\AT#1     {\ensuremath{A_{\mathrm{T}}^{#1}}\xspace}           

\def\C#1      {\ensuremath{\mathcal{C}_{#1}}\xspace}                       
\def\Cp#1     {\ensuremath{\mathcal{C}_{#1}^{'}}\xspace}                    
\def\Ceff#1   {\ensuremath{\mathcal{C}_{#1}^{\mathrm{(eff)}}}\xspace}        
\def\Cpeff#1  {\ensuremath{\mathcal{C}_{#1}^{'\mathrm{(eff)}}}\xspace}       
\def\Ope#1    {\ensuremath{\mathcal{O}_{#1}}\xspace}                       
\def\Opep#1   {\ensuremath{\mathcal{O}_{#1}^{'}}\xspace}                    

\newcommand{\tev}{\ensuremath{\mathrm{\,Te\kern -0.1em V}}\xspace}
\newcommand{\gev}{\ensuremath{\mathrm{\,Ge\kern -0.1em V}}\xspace}
\newcommand{\mev}{\ensuremath{\mathrm{\,Me\kern -0.1em V}}\xspace}
\newcommand{\kev}{\ensuremath{\mathrm{\,ke\kern -0.1em V}}\xspace}
\newcommand{\ev}{\ensuremath{\mathrm{\,e\kern -0.1em V}}\xspace}
\newcommand{\gevc}{\ensuremath{{\mathrm{\,Ge\kern -0.1em V\!/}c}}\xspace}
\newcommand{\mevc}{\ensuremath{{\mathrm{\,Me\kern -0.1em V\!/}c}}\xspace}
\newcommand{\gevcc}{\ensuremath{{\mathrm{\,Ge\kern -0.1em V\!/}c^2}}\xspace}
\newcommand{\gevgevcccc}{\ensuremath{{\mathrm{\,Ge\kern -0.1em V^2\!/}c^4}}\xspace}
\newcommand{\mevcc}{\ensuremath{{\mathrm{\,Me\kern -0.1em V\!/}c^2}}\xspace}

\def\mm   {\ensuremath{\rm \,mm}\xspace}

\def\mum  {\ensuremath{\,\upmu\rm m}\xspace}

\def\mub{\ensuremath{\rm \,\upmu b}\xspace}

\def\invfb   {\ensuremath{\mbox{\,fb}^{-1}}\xspace}

\def\gsim{{~\raise.15em\hbox{$>$}\kern-.85em
          \lower.35em\hbox{$\sim$}~}\xspace}
\def\lsim{{~\raise.15em\hbox{$<$}\kern-.85em
          \lower.35em\hbox{$\sim$}~}\xspace}

\def\pt         {\mbox{$p_{\rm T}$}\xspace}

\def\evtgen     {\mbox{\textsc{EvtGen}}\xspace}
\def\pythia     {\mbox{\textsc{Pythia}}\xspace}

\def\geant      {\mbox{\textsc{Geant4}}\xspace}

\def\tell1  {TELL1\xspace}
\def\ukl1   {UKL1\xspace}

\usepackage{mciteplus}

\begin{document}
\mciteErrorOnUnknownfalse

\renewcommand{\thefootnote}{\fnsymbol{footnote}}
\setcounter{footnote}{1}

\begin{titlepage}
\pagenumbering{roman}

\vspace*{-1.5cm}
\centerline{\large EUROPEAN ORGANIZATION FOR NUCLEAR RESEARCH (CERN)}
\vspace*{1.5cm}
\hspace*{-0.5cm}
\begin{tabular*}{\linewidth}{lc@{\extracolsep{\fill}}r}
\ifthenelse{\boolean{pdflatex}}
{\vspace*{-2.7cm}\mbox{\!\!\!\includegraphics[width=.14\textwidth]{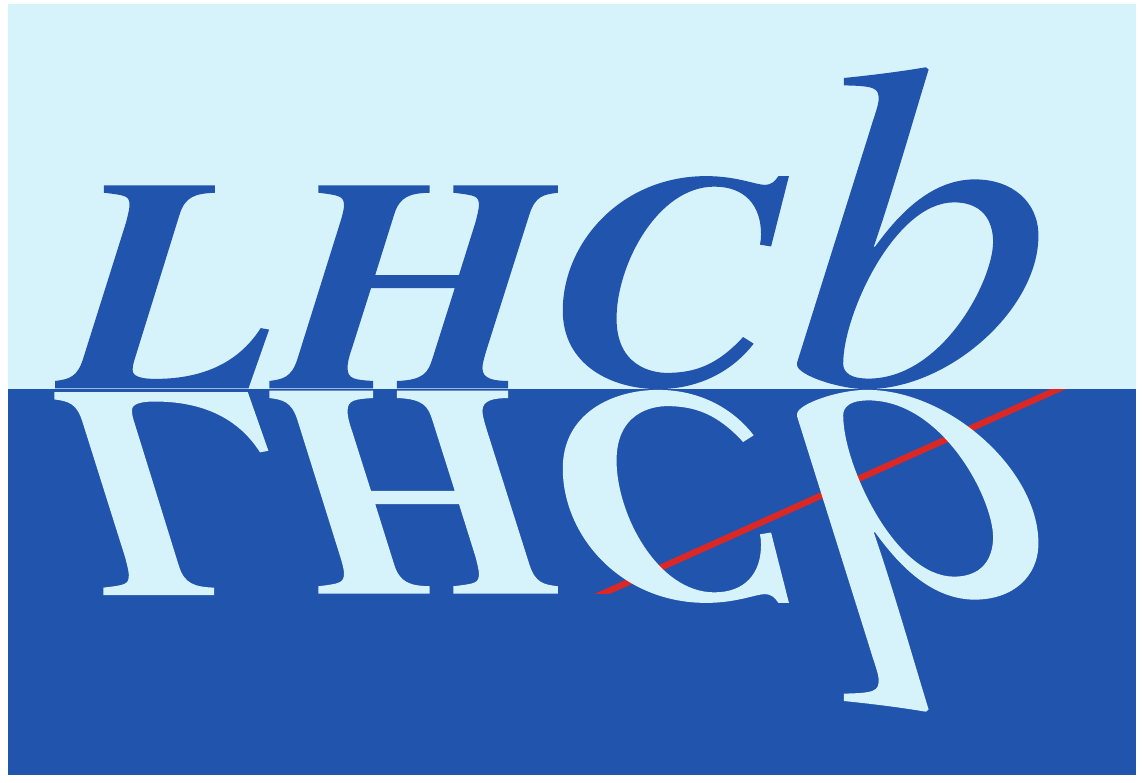}} & &}%
{\vspace*{-1.2cm}\mbox{\!\!\!\includegraphics[width=.12\textwidth]{../figs/lhcb-logo.eps}} & &}%
\\
 & & CERN-PH-EP-2012-305 \\  
 & & LHCb-PAPER-2012-026  \\  
 & &December 19, 2012 \\ 
 & & \\
\end{tabular*}

\vspace*{3.0cm}

{\bf\boldmath\huge
\begin{center}
 Measurement of the \Dpm production asymmetry in 7\tev $pp$ collisions
\end{center}
}

\vspace*{0.5cm}

\begin{center}
The LHCb collaboration\footnote{Authors are listed on the following pages.}
\end{center}

\vspace{\fill}

  \noindent
The asymmetry in the production cross-section $\sigma$ of \Dpm mesons,
\begin{equation*}
A_\mathrm{P} = \frac{\sigma(\Dp) - \sigma(\Dm) }{ \sigma(\Dp) +
\sigma(\Dm) },
\end{equation*}
is measured in bins of pseudorapidity $\eta$ and transverse
momentum \pt within the acceptance of the LHCb detector. The result
is obtained with a sample of $\Dp \to \KS\pip$ decays corresponding to an
integrated luminosity of  1.0\invfb, collected in $pp$ collisions at a centre of mass energy
of 7\tev at the Large Hadron Collider.
When integrated over the kinematic range $2.0< \pt < 18.0$\gevc and $2.20 <
\eta < 4.75$, the production asymmetry is  $A_\mathrm{P} =
(-0.96\pm0.26\pm0.18)\%$. The uncertainties quoted are
statistical and systematic, respectively. The result assumes that any
direct \CP violation in the $\Dp \to \KS\pip$ decay is negligible. No significant dependence on $\eta$ or
\pt is observed.


\vspace*{0.5cm}

\begin{center}
 Published as Physics Letters B \textbf{718} (2013) 902-909
\end{center}

\vspace{\fill}

\end{titlepage}


\clearpage
\setcounter{page}{2}
\mbox{~}
\newpage

\centerline{\large\bf LHCb collaboration}
\begin{flushleft}
\small
R.~Aaij$^{38}$, 
C.~Abellan~Beteta$^{33,n}$, 
A.~Adametz$^{11}$, 
B.~Adeva$^{34}$, 
M.~Adinolfi$^{43}$, 
C.~Adrover$^{6}$, 
A.~Affolder$^{49}$, 
Z.~Ajaltouni$^{5}$, 
J.~Albrecht$^{35}$, 
F.~Alessio$^{35}$
% ...

{\footnotesize \it
% ...
$ ^{5}$Clermont Universit\'{e}, Universit\'{e} Blaise Pascal, CNRS/IN2P3, LPC, Clermont-Ferrand, France\\
$ ^{6}$CPPM, Aix-Marseille Universit\'{e}, CNRS/IN2P3, Marseille, France\\
% ...
$ ^{11}$Physikalisches Institut, Ruprecht-Karls-Universit\"{a}t Heidelberg, Heidelberg, Germany\\
% ...
$ ^{33}$Universitat de Barcelona, Barcelona, Spain\\
$ ^{34}$Universidad de Santiago de Compostela, Santiago de Compostela, Spain\\
$ ^{35}$European Organization for Nuclear Research (CERN), Geneva, Switzerland\\
% ...
$ ^{38}$Nikhef National Institute for Subatomic Physics, Amsterdam, The Netherlands\\
% ...
$ ^{43}$H.H. Wills Physics Laboratory, University of Bristol, Bristol, United Kingdom\\
% ...
$ ^{49}$Oliver Lodge Laboratory, University of Liverpool, Liverpool, United Kingdom\\
% ...
$ ^{n}$LIFAELS, La Salle, Universitat Ramon Llull, Barcelona, Spain\\
% ...
}
\end{flushleft}

\cleardoublepage


\renewcommand{\thefootnote}{\arabic{footnote}}
\setcounter{footnote}{0}

\pagestyle{plain} 
\setcounter{page}{1}
\pagenumbering{arabic}


\section{Introduction}
\label{sec:Introduction}

The Large Hadron Collider (LHC) offers an excellent opportunity to study heavy flavour physics. The rate of production of \ccbar and \bbbar pairs is substantial in the forward region close to the beam direction.
The associated cross-sections were measured at the LHCb experiment in the forward region to be
 $\sigma_{\ccbar} = 1230 \pm 190$\mub 
and $\sigma_{\bbbar} = 74 \pm 14$\mub 
at $\sqrt s = 7$\tev~\cite{LHCb-CONF-2010-013, Aaij:2010gn}.

Direct production of \ccbar pairs at the LHC occurs almost entirely via QCD and electroweak processes that do not discriminate between \cquark and \cquarkbar quarks. However, in hadronization the symmetry is broken by the presence of valence quarks, which introduce several processes that distinguish between \cquark and \cquarkbar quarks~\cite{Norrbin:1999by,Norrbin:2000jy,Norrbin:2000zc}. For example, a \cquark quark could couple to valence quarks to form a charmed baryon, leaving an excess of \cquarkbar quarks. These would hadronize to create an excess of \Dm mesons over \Dp mesons. Furthermore, the kinematic distributions of charmed hadrons and their antiparticles can differ, introducing production asymmetries in local kinematic regions. Analogous production asymmetries in the strange sector are well-established at the LHC, and are seen to be large at high rapidity~\cite{Aaij:2011va}. However, no evidence for a \Dsp production asymmetry was found in a recent study~\cite{LHCb-PAPER-2012-009}.

Searches for \CP violation (CPV) in charmed hadron decays can be used to probe for evidence of physics beyond the Standard Model~\cite{Grossman:2006jg}. Direct CPV is measured using time-integrated observables, and is of particular interest following evidence for CPV in two-body \Dz decays reported by LHCb~\cite{LHCb-PAPER-2011-023} and subsequently by CDF~\cite{Collaboration:2012qw}. In order to understand the origin of this effect, more precise measurements of \CP asymmetries in a suite of decay modes are required. Production asymmetries have the same experimental signature as direct CPV effects and are potentially much larger than the \CP asymmetries to be determined. This problem can sometimes be avoided by taking the difference in asymmetry between two decay modes with a common production asymmetry~\cite{LHCb-PAPER-2011-023} or by studying the difference in kinematic distributions of multi-body decays~\cite{Aaij:2011cw}. However,  these methods result in a reduction in statistical power and are not applicable to all final states. It is therefore important to measure production asymmetries directly.

In this Letter, the \Dpm production asymmetry, defined as
\begin{equation}
  A_{\mathrm{P}} = \frac{\sigma(\Dp)-\sigma(\Dm)}{\sigma(\Dp)+\sigma(\Dm)} ,
  \label{eq:asym}
\end{equation}
for cross sections $\sigma(\Dpm)$, is determined with a sample of $\Dp \to \KS \pi^+$, $\KS \to \pi^+ \pi^-$ decays.\footnote{Charge conjugate decays are implied throughout this letter unless stated otherwise.} As there are no charged kaons in the final state, the detector biases in this decay are simpler to understand than those in other \Dp decays with higher branching fractions. The \KS, a pseudoscalar particle, has a charge-symmetric decay, and the charge asymmetry in the pion efficiency at LHCb has been measured previously for the 2011 data sample~\cite{LHCb-PAPER-2012-009}. However, there is the possibility of CPV in the decay. The expected CPV in the \Dp decay, due to the interference of the Cabibbo-favoured and doubly Cabibbo-suppressed amplitudes, is defined by the charge asymmetry in the partial widths $\Gamma(\Dpm)$,
\begin{equation}
A_{CP} = \frac{\Gamma(\Dp)-\Gamma(\Dm)}{\Gamma(\Dp)+\Gamma(\Dm)}.
\end{equation}
$A_{CP}$ is negligible in the Standard Model: a simple consideration of the CKM matrix leads to a value of at most $1\times10^{-4}$ depending on the strong phase difference between the two amplitudes~\cite{Bigi:1994aw}. Since both amplitudes are at tree level, no enhancement of CPV due to new physics is expected. The current world-best measurement of $A_{CP}$, by the Belle Collaboration, is consistent with zero:
$(0.024 \pm 0.094 \pm 0.067)\%$~\cite{Ko:2012pe, *PhysRevLett.109.119903}. On the other hand, CPV in the neutral kaon system induces an asymmetry which must be considered. This will be discussed further in Sect.~\ref{sec:KSCPV}.

\section{Detector description}
\label{sec:Detector}

The \lhcb detector~\cite{Alves:2008zz} is a single-arm forward
spectrometer covering the \mbox{pseudorapidity} range $2<\eta <5$, designed
for the study of particles containing \bquark or \cquark quarks. The
detector includes a high precision tracking system consisting of a
silicon-strip vertex detector (VELO) surrounding the $pp$ interaction region,
a large-area silicon-strip detector located upstream of a dipole
magnet of reversible polarity with a bending power of about $4{\rm\,Tm}$, and three stations
of silicon-strip detectors and straw drift-tubes placed
downstream. The combined tracking system has a momentum resolution
$\Delta p/p$ that varies from 0.4\% at 5\gevc to 0.6\% at 100\gevc,
and an impact parameter (IP) resolution of 20\mum for tracks with high
transverse momentum \pt. Charged hadrons are identified using two
ring-imaging Cherenkov detectors. Photon, electron and hadron
candidates are identified by a calorimeter system consisting of
scintillating-pad and pre-shower detectors, an electromagnetic
calorimeter and a hadronic calorimeter. Muons are identified by a
system composed of alternating layers of iron and multiwire
proportional chambers. The trigger consists of a hardware stage, based
on information from the calorimeter and muon systems, an
inclusive software stage, which uses the tracking
system, and a second software stage that exploits the full event
information.

\section{Dataset and selection}
\label{sec:sel}

The data sample used in this analysis corresponds to 1.0\invfb of $pp$
collisions taken at a centre of mass energy of $7$\tev at the
Large Hadron Collider in 2011. The polarity of the LHCb
magnetic field was changed several times during the run, and
approximately half of the data were taken with each polarity, referred
to as `magnet-up' and `magnet-down' data hereafter.  To optimise the event selection and estimate
efficiencies, 12.5 million $pp$ collision events containing $\Dp \to \KS\pip$, $\KS \to
\pim\pip$ decays were simulated with \pythia~6.4~\cite{Sjostrand:2006za} with a specific \lhcb
configuration~\cite{LHCb-PROC-2010-056}. Decays of hadronic particles
are described by \evtgen~\cite{Lange:2001uf}. The
interactions of the generated particles with the detector and its
response are implemented using the \geant
toolkit~\cite{Allison:2006ve, *Agostinelli:2002hh} as described in
Ref.~\cite{LHCb-PROC-2011-006}.

Pairs of oppositely
charged tracks with a pion mass hypothesis are combined to form \KS
candidates. Only those \KS candidates with $\pt > 700$\mevc and invariant
mass within 35\mevcc of the nominal value~\cite{PDG2012} are retained. Surviving
candidates are then combined with a third charged track, the bachelor
pion, to form a \Dp
candidate, with the mass of the \KS candidate constrained to its nominal value
in a kinematic fit. Each of the three pion tracks must
be detected in the VELO, so only those \KS mesons that decay well within the VELO
are used. This creates a bias towards short \KS
decay times.
Both the \KS and \Dp candidates are required to have acceptable vertex
fit quality.

\begin{figure}
\begin{center}
\includegraphics*[width=0.67\columnwidth]{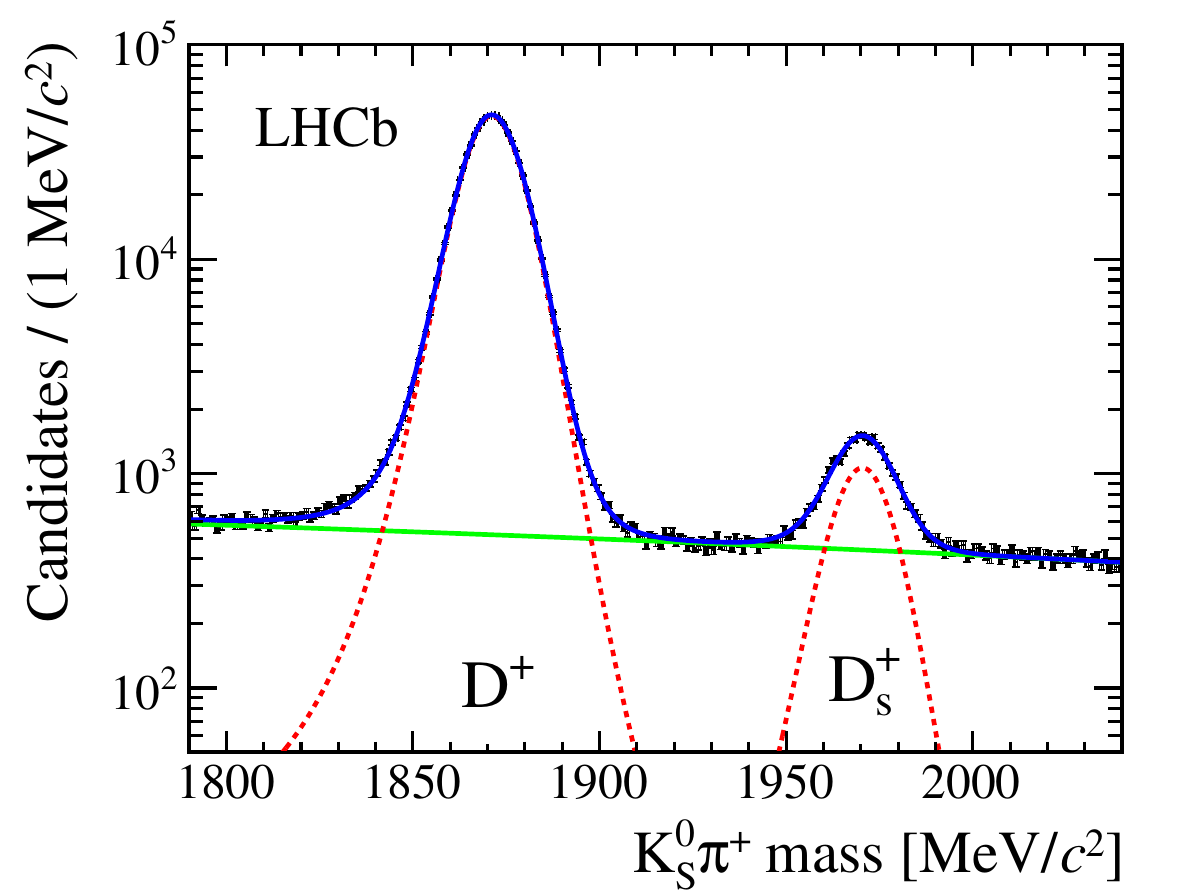}
\end{center}
\caption{
  Mass distribution of selected $\KS\pip$ candidates. The data are
  represented by symbols with error bars. The dashed curves indicate
  the signal and the $\Ds \to \KS\pip$ decays,
  the lower solid line represents the background shape,
  and the upper solid line shows the sum of all fit components.
}
\label{fig:yields}
\end{figure}

Further requirements are applied in order to reduce the
background and to align the selection of bachelor pions with the
dataset used to determine the charge asymmetry in the tracking
efficiency (see Sect.~\ref{sec:results}).
The daughters of the \KS must have $p > 2$\gevc
and $\pt > 250$\mevc. Impact parameter requirements are used to
ensure that both the \KS candidate and its daughter tracks do not originate at any primary vertex (PV) in the event, and
the \KS decay vertex must be at least 10\mm downstream of the PV
with which it is associated.
The bachelor pion must
have $p > 5$\gevc and $\pt > 500$\mevc, be positively
identified as a pion rather than as a kaon, electron or muon, and must
not come from any PV. 
In addition, fiducial requirements are 
applied as in Ref.~\cite{LHCb-PAPER-2011-023} to exclude regions with large 
tracking efficiency asymmetry. 
All three tracks must have an acceptable track fit quality.
The \Dp candidate is required to have $\pt > 1$\gevc,
to point to a PV
(suppressing $D$ from $B$ decays), and to have a decay time
significantly greater than zero. After these
criteria are applied, the remaining background is mostly from random
combinations of tracks. The invariant mass distribution of selected candidates is shown in Fig.~\ref{fig:yields}.

In selected events, a trigger decision may be based on part or all of
the \Dp signal candidate, on other particles in the event, or both. The
second stage of the software trigger is required to find a fully
reconstructed candidate which meets the criteria to be a signal $\Dp
\to \KS\pip$ decay. To control potential charge asymmetries introduced by the
hardware trigger, two possibilities, not mutually exclusive, are
allowed. The hardware trigger decision must be based on one or both of the \KS
daughter tracks, or on a particle other than the decay products of the \Dp candidate.
In both cases, the inclusive software trigger must make a decision based on one of
the three tracks that form the \Dp. For the first case, it is explicitly
required that the same track activated the hardware trigger,
and therefore this is independent of the \Dp charge. The second possibility does not depend
directly on the \Dp charge, but an indirect dependence could be
introduced if the probability for particles produced in association
with the signal candidate to activate the trigger differs between \Dp and
\Dm. This will be discussed further in Sect.~\ref{sec:sys}. After applying
the selection and trigger requirements, 1,031,068 $\KS\pip$ candidates remain.

\section{Yield determination}
\label{sec:yields}

\begin{figure}
\begin{center}
\includegraphics*[width=0.67\columnwidth, viewport=10 10 550
380]{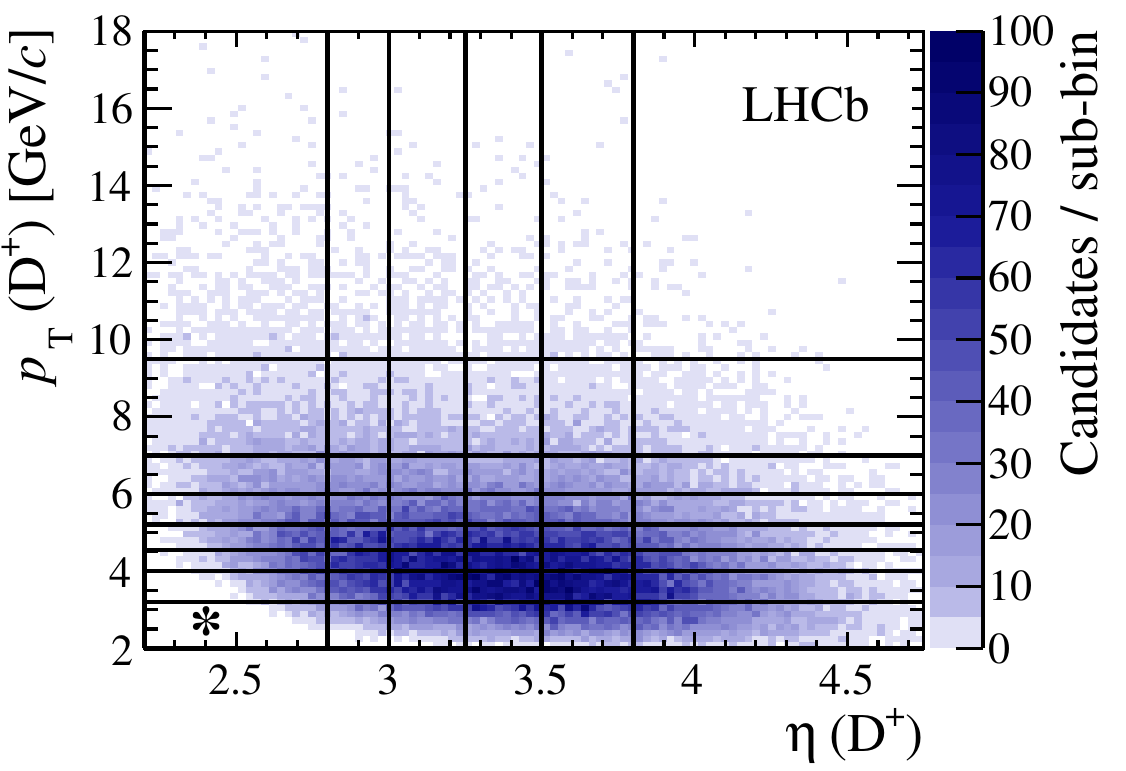}
\end{center}
\caption{
  Background-subtracted distribution of transverse momenta \pt
  versus pseudorapidity $\eta$ for selected $\Dp \to \KS\pip$
  candidates in a signal region of $1845 < m < 1890$\mevcc.
  The bin marked with an asterisk is excluded from the weighted average over the production asymmetries in the bins used to obtain the final result.
}
\label{fig:pteta}
\end{figure}

The signal yields are measured in 48 bins of \pt and $\eta$ using binned
likelihood fits to the distribution of the $\KS\pip$ mass $m$. The bins
are shown in Fig.~\ref{fig:pteta}. The shapes of the $D^{+}_{(s)} \to
\KS\pip$ mass peaks are described by `Cruijff' functions
\cite{delAmoSanchez:2010ae},
\begin{equation}
f(m) \propto \exp\left(\frac{-(m-\mu)^2}{2\sigma_{L, R}^2+(m-\mu)^2\alpha_{L,R}}\right)
\label{eq:cruijff}
\end{equation}
with the measured masses defined by the free parameter $\mu$, the widths by $\sigma_{L}$ and
$\sigma_{R}$, and the tails by $\alpha_{L}$ and $\alpha_{R}$. The parameters
$\alpha_{L}$ and $\sigma_{L}$ are used for $m < \mu$ and
$\alpha_{R}$ and $\sigma_{R}$ for $m > \mu$. The background is fitted
with a straight line plus an additional Gaussian component to account
for background from $\Dsp \to \KS\pip\piz$ decays. The yield of the latter is
consistent with zero in most \pt, $\eta$ bins. The fit is performed
simultaneously over four subsamples (\Dp magnet-up, \Dp magnet-down,
\Dm magnet-up, and \Dm magnet-down data) with the masses and yields of
the $D^{\pm}_{(s)}$, and the yield of background, allowed to vary independently
in the four subsamples. All other parameters are shared. The charge
asymmetries are then determined from the yields. The results are cross-checked with a sideband subtraction procedure under the assumption of a linear background.

\section{Effect of \emph{CP} violation in the neutral kaon system}
\label{sec:KSCPV}

\begin{figure}
\begin{center}
\includegraphics*[width=0.67\columnwidth,viewport=5 0 545
425]{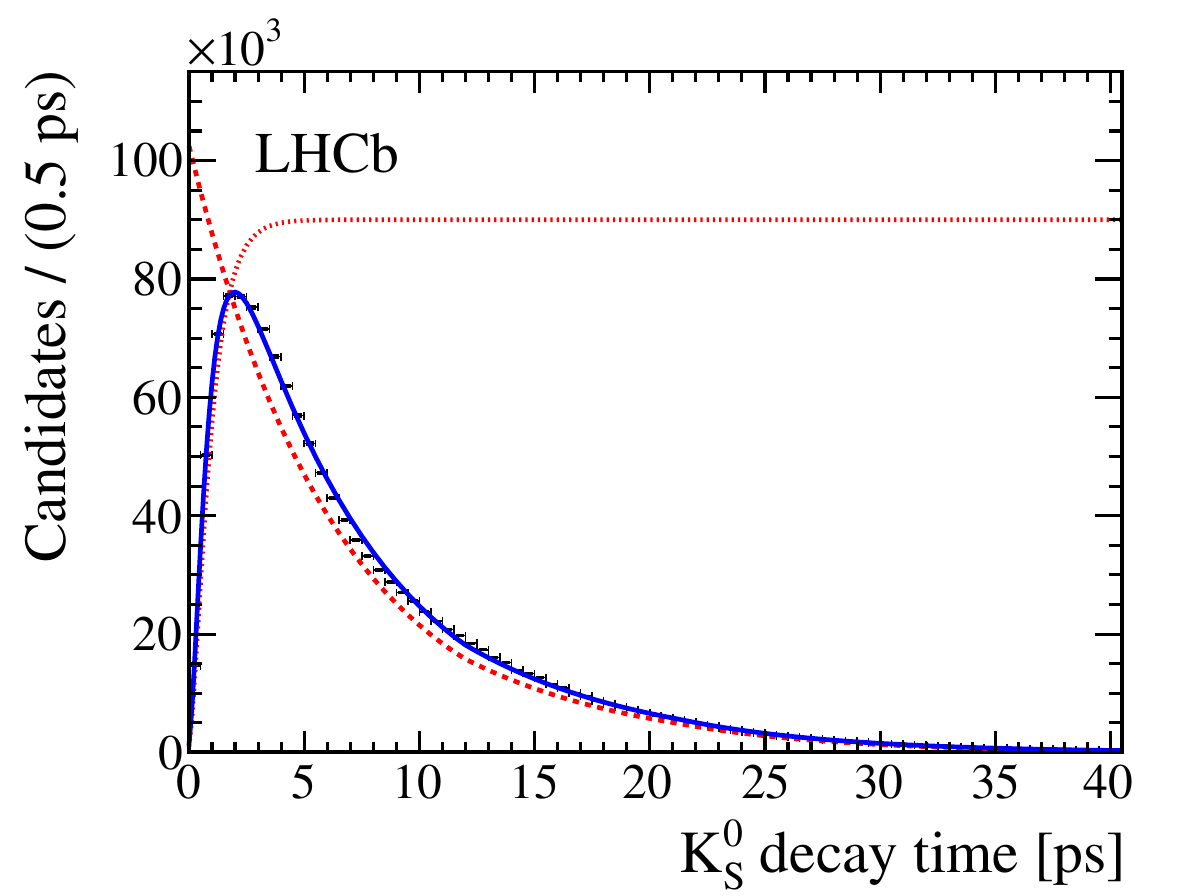}
\end{center}
\caption{
  Observed \KS decay time distribution within the LHCb acceptance. The data points are fitted with an empirical
  function (solid curve). This contains a component for the upper
  decay time acceptance, due mainly to the requirement that the \KS decays inside the VELO (dashed curve) and
  a component for the lower decay time acceptance, due to the selection cuts (dotted curve).
  These are shown scaled by arbitrary factors. The CPV is not sensitive to the fine
  details of the distribution, so the fit quality is not important.
}
\label{fig:kslt}
\end{figure}

\CP violation in the neutral kaon system can affect the observed
asymmetry in the $\Dp \to \KS \pip$ decay~\cite{Bianco:2003vb}.
The bias on $A_{\mathrm{P}}$ due to the CPV depends on the decay time
acceptance $F(t)$ of the \KS meson, according to 
\begin{equation}
A_{\epsilon} \sim 2\Re(\epsilon)\left[1-\frac{\int_{0}^{\infty} F(t) e^{-\frac{1}{2}(\Gamma_{\mathrm{S}} + \Gamma_{\mathrm{L}}) t}\left(\cos \Delta mt - \frac{\Im(\epsilon)}{\Re(\epsilon)}\sin\Delta mt\right)dt}{\int_{0}^{\infty} F(t) e^{-\Gamma_{\mathrm{S}} t}dt}\right]
,
\label{eq:KsCPV:grossman2}
\end{equation}
where $\epsilon$ parameterises the indirect CPV in neutral kaon
mixing, $\Gamma_{\mathrm{S}}$ and $\Gamma_{\mathrm{L}}$ are the decay widths of the
\KS and \KL respectively, and $\Delta m$ is their mass difference~\cite{Grossman:2011zk, 2000EPJC...18...41C}. Direct CPV and terms of order $\epsilon^{2}$ are neglected. To
determine the decay time acceptance, the \KS
decay time is fitted with an empirical function shown in
Fig.~\ref{fig:kslt}. All of the \KS candidates used in this
analysis decay inside the VELO with an average measured
lifetime of $6.97\pm0.02$~ps, which is much shorter
than the nominal \KS lifetime of 89.5~ps. Using $\Re(\epsilon) = 1.65\times10^{-3}$~\cite{PDG2012} in
Eq.~\ref{eq:KsCPV:grossman2}, we obtain $A_{\epsilon}  =
(2.831^{+0.003}_{-0.004})\times10^{-4}$ for the CPV in the neutral
kaon system, where the uncertainty quoted is statistical only. This
value is subtracted from the measured production asymmetry and a systematic uncertainty equal to its central value is assigned. 

\section{Results}
\label{sec:results}

In order to convert the measured charge asymmetries in the 48 bins of
\pt and $\eta$ into production asymmetries, a correction for the
asymmetry in the pion reconstruction efficiency is made.
This asymmetry was evaluated previously in eight bins of pion azimuthal angle
$\phi$ and two bins of pion momentum with a control sample of $\Dstarp
\to \Dz \pi^+$, $\Dz \to \Km\pip\pim\pip$ decays in the same dataset~\cite{LHCb-PAPER-2012-009}. The average efficiency asymmetry ratios $\epsilon_{\pip}/\epsilon_{\pim}$ in that sample were found to be $0.9914\pm0.0040$  for magnet-up data and $1.0045\pm0.0034$ for magnet-down data.

After the correction is applied, the resulting asymmetries for magnet-up and magnet-down data in
each \Dp \pt and $\eta$ bin are averaged with equal weights to obtain
the production asymmetries in two-dimensional bins of \pt and $\eta$, given in Table~\ref{tab:binnedAP}.
Any left-right asymmetries that differ between the signal $\Dp \to
\KS\pip$ decay and the $\Dz \to \Km\pip\pim\pip$ control channel will cancel in this average.

\begin{figure}
\includegraphics*[width=0.47\textwidth,viewport=5 0 540 400]{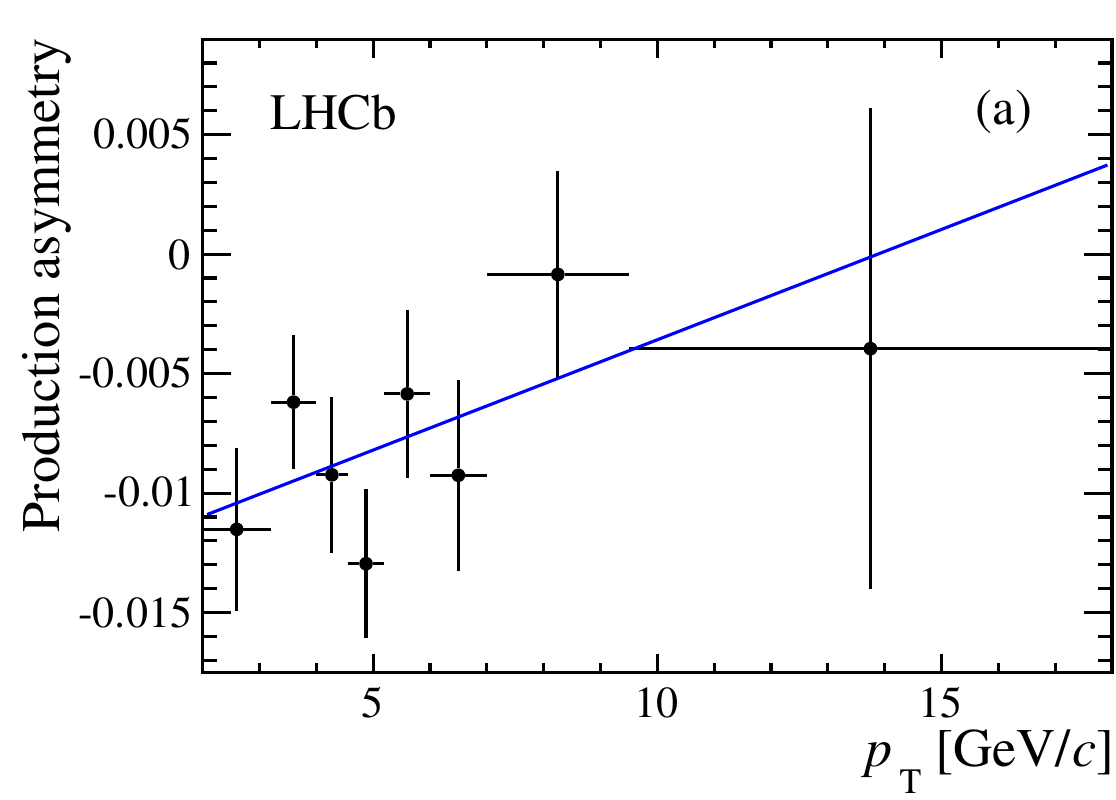}
\includegraphics*[width=0.47\textwidth,viewport=5 0 540
400]{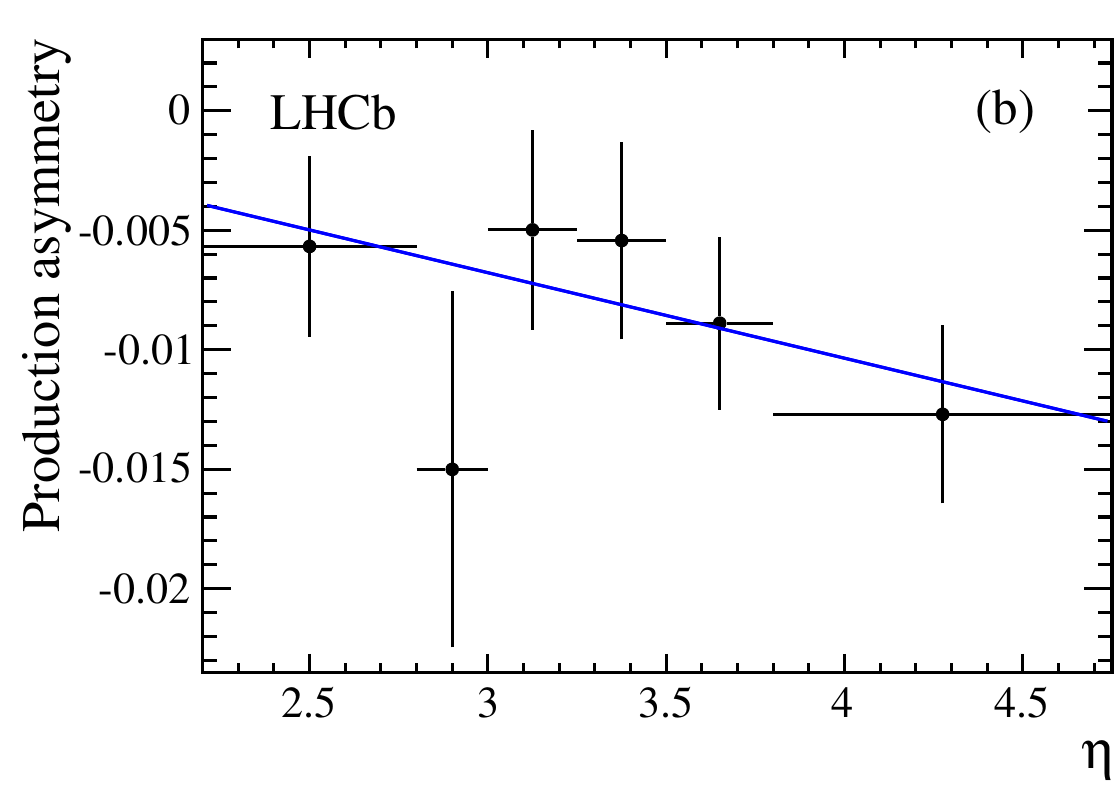}

\caption{Production asymmetry as a function of (a) transverse momentum
  \pt and (b) pseudorapidity $\eta$. The straight line fits have slopes
 of $(0.09\pm0.07) \times10^{-2}$ (\gevc)$^{-1}$ and $(-0.36\pm0.28)$\%, and values of
  $\chi^{2}$ per degree of freedom of $5.5/6$ and $2.2/4$, respectively. The
 error bars include only the statistical uncertainty on the \Dp signal sample and are uncorrelated within a
 given plot.}
\label{fig:prodpteta}
\end{figure}

Reconstruction and selection efficiencies from the simulation are
used to calculate binned efficiency-corrected yields. These are used
to weight the production asymmetries in the average over the \pt
and $\eta$ bins. The result is an asymmetry for \Dp produced in the
LHCb acceptance. The same weighting technique is applied to obtain
production asymmetries as one-dimensional functions of \pt and
$\eta$. The bin marked with an asterisk in Fig.~\ref{fig:pteta} has a
high cross section but 
is mostly outside the acceptance and so it is
excluded from the average.
After subtracting the contribution from CPV in the kaon system, the
production asymmetry is $(-0.96\pm0.19\pm0.18)$\%. The uncertainties are the statistical errors on the $\Dp \to \KS\pip$ yields
and that due to the tagged $\Dz \to
\Km\pip\pim\pip$ sample used to calculate the pion efficiencies. Summing
these in quadrature, we obtain
\begin{equation*}
  A_{\mathrm{P}} = \left( -0.96 \pm 0.26 \, (\mathrm{stat.}) \right)\% .
  \label{eq:resultStatErr}
\end{equation*}
The production asymmetry as a function of \pt and $\eta$ is
given in Fig.~\ref{fig:prodpteta}. No significant dependence of the
asymmetry on these variables is observed.
As a cross-check, the average production asymmetry is calculated for
magnet-up and magnet-down data separately, and found to be fully
consistent: $(-1.07\pm0.41)\%$ and $(-0.85\pm0.34)\%$,
respectively.

\begin{table*}
  \caption{
    Production asymmetry for \Dp mesons, in percent, in $(\pt, \eta)$ bins, for $2.0 < \pt < 18.0$\gevc and $2.20 < \eta < 4.75$.
    The uncertainties shown are statistical only; the systematic uncertainty is $0.17\%$ (see Table~\ref{tab:sys:summary}).
  }
\begin{adjustwidth}{-.5in}{-.5in}
  \begin{center}
\small
    \begin{tabular}{c|cccccc}
      & \multicolumn{6}{c}{$\eta$} \\
      \pt (\gevc) & $(2.20,2.80)$ & $(2.80,3.00)$ & $(3.00,3.25)$ & $(3.25,3.50)$ & $(3.50,3.80)$ & $(3.80,4.75)$ \\ \hline
$(2.00,3.20)$ & $-0.0 \pm 2.5  $ & $-2.2 \pm 1.2  $ & $-0.4 \pm 0.8  $ & $-0.4 \pm 0.7  $ & $-1.2 \pm 0.6  $ & $-1.2 \pm 0.5  $ \\
$(3.20,4.00)$ & $-0.4 \pm 0.9  $ & $-0.4 \pm 0.7  $ & $-0.4 \pm 0.5  $ & $-1.1 \pm 0.5  $ & $+0.1 \pm 0.5  $ & $-1.2 \pm 0.5  $ \\
$(4.00,4.55)$ & $+0.1 \pm 0.8  $ & $-1.0 \pm 0.8  $ & $-1.3 \pm 0.6  $ & $-2.0 \pm 0.6  $ & $-0.1 \pm 0.6  $ & $-2.1 \pm 0.7  $ \\
$(4.55,5.20)$ & $-1.6 \pm 0.7  $ & $-0.6 \pm 0.8  $ & $-0.5 \pm 0.6  $ & $-0.7 \pm 0.6  $ & $-1.6 \pm 0.6  $ & $-2.0 \pm 0.8  $ \\
$(5.20,6.00)$ & $-0.5 \pm 0.7  $ & $-0.8 \pm 0.8  $ & $+0.2 \pm 0.7  $ & $-0.3 \pm 0.7  $ & $-0.6 \pm 0.7  $ & $-1.2 \pm 0.9  $ \\
$(6.00,7.00)$ & $-1.4 \pm 0.8  $ & $+0.5 \pm 1.0  $ & $-0.9 \pm 0.9  $ & $-0.6 \pm 0.9  $ & $-0.7 \pm 0.9  $ & $-1.6 \pm 1.2  $ \\
$(7.00,9.50)$ & $-0.4 \pm 0.8  $ & $-0.4 \pm 1.1  $ & $-0.2 \pm 1.1  $ & $+1.7 \pm 1.1  $ & $-1.4 \pm 1.1  $ & $+1.2 \pm 1.4  $ \\
$(9.50,18.00)$ & $-0.6 \pm 1.3  $ & $+1.8 \pm 2.3  $ & $-2.5 \pm 2.2  $ & $+1.8 \pm 2.4  $ & $+1.1 \pm 2.5  $ & $\phantom{0}\mathord{-}7 \pm 11  $ \\

    \end{tabular}
  \end{center}
\end{adjustwidth}
  \label{tab:binnedAP}
\end{table*}

\section{Systematic uncertainties}
\label{sec:sys}

The sources of systematic uncertainty are summarised in Table \ref{tab:sys:summary}. The dominant uncertainty of $1.5\times10^{-3}$ is due to asymmetries introduced by the trigger. Events which are triggered independently of the signal decay, i.e. by a track that does not form part of the signal candidate, could be triggered by particles produced in association with the \Dp meson. If this occurs, the asymmetry in this sample would be correlated with the production asymmetry, and would bias the measurement of it. This was studied with a control sample of the abundant $\Dp \to \Km\pip\pip$ decay.
To mimic the charge-unbiased sample of $\Dp \to \KS \pi^+$ decays which are triggered by a \KS
daughter, we choose the kaon and one pion at random and require that the trigger decision is based on one of these tracks. This is close to being charge-symmetric between \Dp and \Dm candidates, with some residual
effects due to differences in material interaction between $K^+$ and $K^-$ mesons. The raw asymmetry in this subsample of $\Dp \to \Km\pip\pip$ decays is then compared to that in the much larger sample of candidates that are triggered independently of the signal decay. The difference in raw charge asymmetry between these two samples, $(1.5\pm0.4)\times10^{-3}$, is a measure of the scale of the bias. Unlike the signal, the $\Km\pip\pip$ decay also includes a component due to the $K^+/K^-$ asymmetry, and therefore this is treated as a systematic uncertainty rather than a correction. This is cross checked with other control samples such as $\Dsp \to \phi\pip$ and the uncertainty is found to be conservative.

\begin{table}
  \caption{Summary of absolute values of systematic uncertainties on $A_{\mathrm{P}}$.
    For the binned production asymmetries given in Table~\ref{tab:binnedAP},
    all uncertainties except that on the reconstruction efficiency apply,
    giving a combined systematic uncertainty of 0.17\%.
  }
\begin{center}
\begin{tabular}{lc}
Systematic effect & Uncertainty (\%) \\
\hline
Trigger asymmetries & 0.15 \\
$D$ from $B$ & 0.04 \\
Selection criteria & 0.05 \\
Running conditions & 0.04 \\
Pion efficiency & 0.02 \\
Fitting & 0.04 \\
Kaon \CP violation &0.03 \\
Weights (reconstruction efficiency) & 0.05 \\
\hline
Total including uncertainty on weights & 0.18 \\
\end{tabular}
\end{center}
\label{tab:sys:summary}
\end{table}

Further systematic uncertainties arise from the contamination of the prompt sample by $D$ candidates that originate from $B$ decays. The yield of these is calculated using the measured cross-sections~\cite{LHCb-CONF-2010-013,Aaij:2010gn}, branching ratios, and efficiencies determined from the simulation. The fraction of $D$ candidates from $B$ decays is found to be $(1.2\pm0.3)\%$. This quantity is combined with the \Bz production asymmetry, which is estimated to be $(-1.0\pm1.3)\%$~\cite{LHCb-PAPER-2011-029}, to determine the systematic uncertainty. 

Certain selection criteria differ between the $\Dp \to \KS\pip$ signal sample and the $\Dz \to \Km\pip\pim\pip$ decays used to determine the asymmetry in the pion efficiencies. The charge asymmetry is found to depend weakly on the value of the requirement on the pion \pt. Pions in the signal sample must have \pt $> 500$\mevc while those in the control sample must have \pt $> 300$\mevc. A systematic uncertainty is calculated by estimating the proportion of signal candidates with $300 < \pt < 500$\mevc and multiplying this fraction by the difference between the charge asymmetries in the low \pt region and the average.

The difference in signal yields per pb$^{-1}$ of integrated luminosity between magnet-up and magnet-down data is used to determine a systematic uncertainty for changes in running conditions that could impair the cancellation of detector asymmetries achieved by averaging over the magnet polarities. There is also a systematic uncertainty on the pion efficiency asymmetry associated with the determination of the yields of $\Dz \to\Km\pip\pim\pip$ decays.  The error associated with the mass fit is determined by comparing fitted and sideband-subtracted results. The CPV in the neutral kaon decay, discussed in Sect.~\ref{sec:KSCPV}, is also included as a systematic uncertainty.

Other systematic effects such as regeneration in the neutral kaon system~\cite{PhysRevLett.38.1116}, second order effects due to the kinematic binning of the $\Dp \to \KS\pip$ sample, and asymmetric backgrounds such as that from $\Dsp \to \KS\Kp$ with the kaon misidentified as a pion, were considered but found to be negligible. When taking the average asymmetry weighted by the efficiency-corrected yield in each bin, the limited number of simulated events leads to an uncertainty on the reconstruction efficiency and hence on the per-bin weights. This does not contribute to the uncertainty on the individual asymmetries given in Table~\ref{tab:binnedAP}, which are calculated without using the simulation. 
A quadratic sum yields an overall systematic uncertainty of $1.8\times10^{-3}$. 

In principle, CPV in the charm decay could occur via the interference of Cabibbo-favoured and doubly Cabibbo-suppressed amplitudes, but this is strongly suppressed by the CKM matrix and no evidence for it has been observed at the $B$-factories~\cite{delAmoSanchez:2011zza, Ko:2012pe}. If we allowed for the possibility of new physics or large unexpected enhancements of the Standard Model CPV in these tree-level \Dp decays, the uncertainty on the null result found at Belle \cite{Ko:2012pe} would increase the total systematic uncertainty to $2.1\times10^{-3}$.

\section{Conclusions}
\label{sec:conc}

Evidence for a charge asymmetry in the production of \Dp
decays is observed at LHCb. In
the kinematic range $2.0 < \pt < 18.0$\gevc and $2.20 < \eta < 4.75$,
excluding the region with $2.0 < \pt < 3.2$\gevc, $2.20<\eta <2.80$, the
average asymmetry is 
\begin{equation*}
A_{\mathrm{P}}= (-0.96\pm0.26\pm0.18)\%,
\end{equation*}
where the first uncertainty is statistical and the second is systematic. The result is inconsistent with zero at approximately three standard deviations. There is no evidence
for a significant dependence on \pt or pseudorapidity at the present level of precision.
The bias on the measured asymmetry due to \CP
violation in kaon decays has been calculated and found to be almost negligible
for this dataset. These results are consistent with
expectations~\cite{Norrbin:2000zc}
and lay the foundations for searches for \CP
violation in Cabibbo suppressed \Dp decays.

\section*{Acknowledgements}

\noindent We express our gratitude to our colleagues in the CERN
accelerator departments for the excellent performance of the LHC. We
thank the technical and administrative staff at the LHCb
institutes. We acknowledge support from CERN and from the national
agencies: CAPES, CNPq, FAPERJ and FINEP (Brazil); NSFC (China);
CNRS/IN2P3 and Region Auvergne (France); BMBF, DFG, HGF and MPG
(Germany); SFI (Ireland); INFN (Italy); FOM and NWO (The Netherlands);
SCSR (Poland); ANCS/IFA (Romania); MinES, Rosatom, RFBR and NRC
``Kurchatov Institute'' (Russia); MinECo, XuntaGal and GENCAT (Spain);
SNSF and SER (Switzerland); NAS Ukraine (Ukraine); STFC (United
Kingdom); NSF (USA). We also acknowledge the support received from the
ERC under FP7. The Tier1 computing centres are supported by IN2P3
(France), KIT and BMBF (Germany), INFN (Italy), NWO and SURF (The
Netherlands), CIEMAT, IFAE and UAB (Spain), GridPP (United
Kingdom). We are thankful for the computing resources put at our
disposal by Yandex LLC (Russia), as well as to the communities behind
the multiple open source software packages that we depend on.

\addcontentsline{toc}{section}{References}
\bibliographystyle{LHCb}
\bibliography{main}

\end{document}